\documentclass[twocolumn,aps,prl,showpacs]{revtex4-1}
\usepackage{dcolumn}
\usepackage{bm}
\usepackage{amsmath}
\usepackage{amssymb}
\usepackage{revsymb}
\usepackage{graphics}
\usepackage{graphicx}

\begin{document}
\title{Entropy density and entropy flux in near-field thermal radiation}
\author{Arvind Narayanaswamy}
\email{arvind.narayanaswamy@columbia.edu}
\author{Yi Zheng}
\affiliation{%
Department of Mechanical Engineering, Columbia University\\
New York, NY 10027
}

\begin{abstract}
We propose a method to evaluate the entropy density and entropy flux in a vacuum gap between two half-spaces that takes into account influence of near-field effects, i.e., interference, diffraction, and tunneling of waves. The method developed is used to determine the maximum work that can be extracted through near-field radiative transfer between two half-spaces at different temperatures.
\end{abstract}
\maketitle


The inception of modern quantum physics can be traced back to Planck's pioneering work on blackbody radiation \cite{planck}. Central to Planck's work is the thermodynamic analysis of thermal radiation in a cavity, which requires knowledge of energy, momentum, and entropy of photons. Planck's analysis is restricted to the case when near-field effects (NFE), i.e., the collective influence of diffraction, interference, and  tunneling of waves, are absent \cite{planck}.
 Though we know much about energy (leading to radiative heat transfer) and momentum (leading to van der Waals and Casimir forces) associated with near field radiative transfer (NFRT), little, if any, is known about entropy \cite{zhang2007entropy}. 
   The purpose of this letter is to determine entropy density and entropy flux in a planar vacuum cavity between two half-spaces at different temperatures when the separation between them is small enough that NFE are important. Using the entropy flux, the maximum work that can be extracted through NFRT between two half-spaces at different temperatures is determined.


Planck showed that the far-field radiation entropy intensity is given by \cite{planck}: 
\begin{equation}
\label{eqn:planckentropyintensity}
l_{\omega}(\hat{\Omega})= k_B\frac{ \beta \omega^2}{8\pi^3 c^2}[ (1+ M_{\omega}) \ln (1+ M_{\omega} ) -   M_{\omega} \ln M_{\omega}  ]  
\end{equation}
where $ k_B $ is the Boltzmann constant, $\omega$ is the angular frequency, $ c $ is the speed of light in free space, $\displaystyle M_{\omega} \equiv M(\omega,\hat{\Omega}) = \frac{I(\omega,\hat{\Omega})}{\beta \hbar\omega^3/8 \pi^3 c^3} $, $ 2\pi\hbar $ is the Planck constant, and $ I(\omega,\hat{\Omega}) $ is the spectral radiation intensity of a ray in the direction defined by unit vector $ \hat{\Omega} $. $ \beta = 1 $ for polarized radiation and $ \beta = 2 $ for unpolarized radiation.  The spectral entropy density is given by $ s_{\omega} = (1/c)\int l_{\omega}(\hat{\Omega}) d\Omega $. von Laue \cite{laue1906thermodynamik,*nigam1997laue} and other researchers  \cite{barakat1993neumann,*rueda1973entropy} investigated the effect of partial coherence on the entropy of  radiation. 
  Petela and others \cite{petela1964exergy,*parrott1978theoretical,landsberg80a} used the concept of availability or exergy to determine the maximum work that can be extracted from solar radiation. All of the above mentioned works \cite{laue1906thermodynamik,nigam1997laue,barakat1993neumann,rueda1973entropy,petela1964exergy,parrott1978theoretical,landsberg80a} rely on Eq. \ref{eqn:planckentropyintensity} for entropy intensity. Underlying the derivation of Eq. \ref{eqn:planckentropyintensity} is the assumption that the electromagnetic local density of states (LDOS) and velocity of radiation are independent of position and given by their values in free space. The implications of the breakdown of the above-mentioned assumptions when NFE are present, to the best of our knowledge, have not been investigated. 
  
\begin{figure}[b]
\centering
\includegraphics[width=3.in]{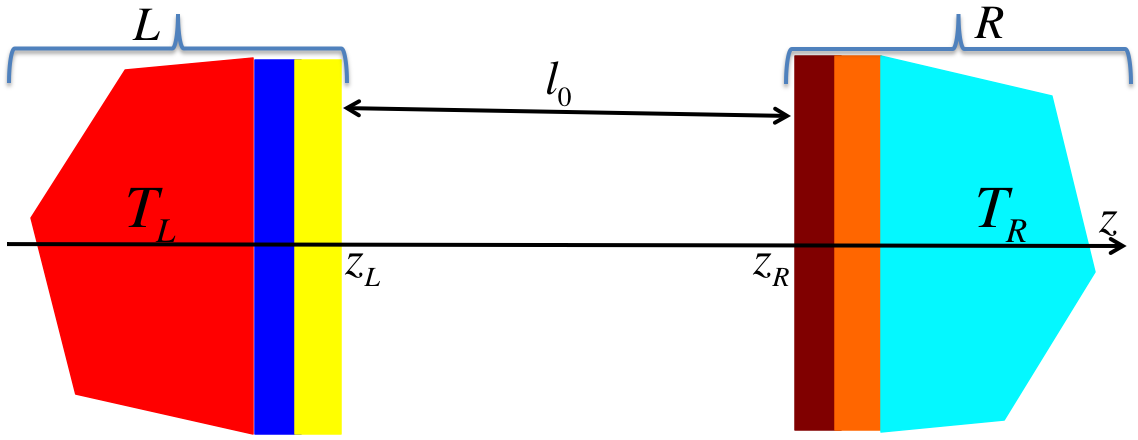}
\caption{\label{fig:config}(Color online) Two half-spaces $ L $ and $ R $ separated by a vacuum gap of thickness $ l_0 $.} 
\end{figure}

Consider a planar vacuum cavity of thickness $ l_0 = |z_R-z_L| $ between two parallel half-spaces $ L $ and $ R $, as shown in Fig. \ref{fig:config}, at temperatures $ T_L $ and $ T_R $ respectively. Each half-space is assumed to be isothermal.
 Half-spaces with planar thin films do not introduce any conceptual difficulties and are taken into account via generalized Fresnel reflection coefficients $\tilde{R}_{0L}$ and $\tilde{R}_{0R}$. Because of translational symmetry in the $ xy $ plane, all quantities pertinent to this letter only depend on $z$.
The influence of NFE is captured by modeling thermal radiation using Rytov's theory of fluctuational electrodynamics (FE) \cite{rytov59a}. In FE, the cross power spectral densities of the fluctuating current sources are related to the electromagnetic properties and temperature through the fluctuation-dissipation theorem \cite{eckhardt1982first}. 
 The ensemble averaged Poynting vector and energy density can be written in terms of the dyadic Green's functions (DGFs) of the vector Helmholtz equation, which are well known for layered media \cite{chew95a}. Though fluctuational electrodynamics provides us with a framework for determining energy density and Poynting vector, it does not tell us how to compute entropy.

Entropy can be calculated by determining the number of microscopic states into which a given number of photons can be distributed. Consider a thin film of area $ A $, which can be made as large as necessary, lying between coordinates $ z$ and $z + dz $ in the vacuum gap. We first consider the case when $ T_L \neq T_R $. 
The entropy of electromagnetic waves emitted from half-space $ L $ with microscopic states in the range $ \mu $ to $\mu + d\mu $ within this thin film is given by $ s_L(\mu,z)d\mu Adz$ and obtained from the relation: $ s_L(\mu,z)d\mu Adz = k_B \ln \delta W_L $. Here $\mu$ is the set of variables defining the space containing the  microscopic states, which is unidentified as yet, and $ \delta W_L $ is the number of ways in which $ n_L(\mu, z)d\mu Adz$ photons can be distributed across the number of accessible microscopic states, represented by $ \rho_a(\mu,z) d\mu Adz$. $n_L(\mu, z)$ and $ \rho_a(\mu,z) $ are the number density of photons and accessible microscopic states per unit volume in the $\mu$ space. The total number of available microscopic states is related to the LDOS $\rho (\omega,z) $ and is given by $\rho(\mu,z)d\mu Adz$, of which only a fraction $\rho_a(\mu,z)/\rho(\mu,z)$  is accessible to the photons from $L$. $\delta W_L $ is given by \cite{tolman1938principles} 
$ \frac{(n_L(\mu,z)d\mu Adz + \rho_a(\mu,z)d\mu Adz -1)!}{ (n_L(\mu,z)d\mu Adz)! (\rho_a(\mu,z)d\mu Adz -1)! } $. A similar expression exists for $\delta W_R $. Taking the two independent polarizations into account, we can write (assuming that $ \rho_a(\mu,z)d\mu Adz \gg 1 $):
\begin{equation}
\label{eqn:genentropydensityright}
\begin{split}
s^{(j)}_h(\mu,z)=  & k_B \rho_a^{(j)}(\mu,z) \Big[- \frac{n^{(j)}_h(\mu,z)}{\rho_a^{(j)}(\mu,z) } \ln \Big[\frac{n^{(j)}_h(\mu,z)}{\rho_a^{(j)}(\mu,z) } \Big] \\
&+ \Big(1+\frac{n^{(j)}_h(\mu,z)}{\rho_a^{(j)}(\mu,z) } \Big) \ln \Big(1+\frac{n^{(j)}_h(\mu,z)}{\rho_a^{(j)}(\mu,z) } \Big) \Big]
\end{split}
\end{equation}
where $ h=L,R $ refers to contributions from half-spaces $ L $ and $ R $ respectively, and $ j=s,p $ refers to $ s $ or $ p $ polarization. 
Since the electromagnetic waves emitted from the two half-spaces are incoherent, the total entropy density at any location is given by $ s(z) = \int d\mu \sum\limits_{j=s,p} ( s^{(j)}_L(\mu,z) + s^{(j)}_R (\mu,z ) )  $ \cite{barakat1983nfold}. 

$ n^{(j)}_h(\mu, z)  $ can be determined from its relation to $ u_h^{(j)}(z) $, the energy density in the vacuum gap of $ s $ or $ p $ polarized waves from half-space $ h $. The method used by Antezza et. al. \cite{antezza2008casimir} can be used to determine $ u_h^{(j)}(z) $ \footnote{The authors of Ref. \cite{antezza2008casimir} evaluate the $zz$ component of the stress tensor. Minor changes have to be made to their method to calculate the energy density.}.  $ n^{(j)}_h(\mu, z)  $ and $ u_h^{(j)}(z) $ are related as follows:
\begin{equation}
\label{eqn:energydensity} u^{(j)}_{L}(z) = \int_{PW} d\mu \hbar\omega n^{(j)}_L(\mu, z) + \int_{EW} d\mu \hbar\omega n^{(j)}_L(\mu, z)
\end{equation}
where $ PW (EW)$ refers to propagating (evanescent) waves. The integrals $\int_{PW}$ and $\int_{EW}$ will be defined shortly. $ n^{(j)}_L(\mu, z)$  is given by (Eq. \ref{eqn:njLPW} for PW and Eq. \ref{eqn:njLEW} for EW):
\begin{subequations}
\label{eqn:njL}
\begin{equation}
\label{eqn:njLPW}
\begin{split}
 n^{(j)}_{L}(\mu,z) =& \frac{(1-|\tilde{R}^{(j)}_{0L}|^2)}{8\pi^3 [\exp(\hbar\omega/k_BT_L)-1]} \times\\
& 
\displaystyle \frac{[1+|\tilde{R}^{(j)}_{0R}|^2+\frac{2k_{\rho}^2}{k_0^2}\Re( \tilde{R}^{(j)}_{0R} e^{i2k_{z0} (z_R-z)}) ]}{ \big|1-\tilde{R}^{(j)}_{0L}\tilde{R}^{(j)}_{0R} e^{i2k_{z0}l_0} \big|^2} 
\end{split}
\end{equation}
\begin{equation}
\label{eqn:njLEW}
\begin{split}
& n^{(j)}_{L}(\mu,z) = \frac{4\Im(\tilde{R}^{(j)}_{0L})  e^{-2\beta_{z0}l_0}}{8\pi^3 [\exp(\hbar\omega/k_BT_L)-1]}  \times\\
& 
 \frac{   [
\Re( \tilde{R}^{(j)}_{0R}) + \frac{k_{\rho}^2}{2k_0^2} (e^{2\beta_{z0} (z_R-z)}+
 |\tilde{R}^{(j)}_{0R}|^2  e^{-2\beta_{z0} (z_R-z)} )]}{ \big|1-\tilde{R}^{(j)}_{0L}\tilde{R}^{(j)}_{0R} e^{-2\beta_{z0}l_0} \big|^2\sqrt{2\frac{k_{\rho}^2}{k_0^2}-1}}  
\end{split}
\end{equation}
\end{subequations}
where $k_{\rho}$ is the magnitude of the in-plane wave vector $ {\mathbf{k}}_{\rho} = k_{\rho} \mathbf{\hat{k}}_{\rho} = k_x \hat{x} + k_y \hat{y} $, $ k_0=\omega/c $, $ k_{\rho}^2+k_{z0}^2 = k_0^2 $ for $ PW $ ($ k_{\rho} < k_0 $), and $ k_{\rho}^2-\beta_{z0}^2 = k_0^2 $ for $ EW $ ($ k_{\rho} > k_0 $). 

\begin{figure}
\centering
\includegraphics[width=8.5cm]{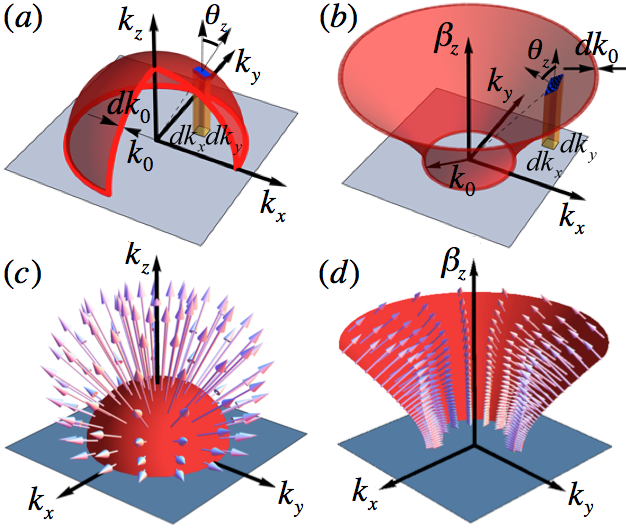}
\caption{\label{fig:phasespacepwew} (Color online) Constant frequency surfaces for (a)  $PW$, and (b) $EW$. The surfaces are symmetric about the $ k_x-k_y $ plane.  In (c) and (d), the arrows are proportional to and point in the direction of the generalized radiation intensity for (c) $PW$, and (d) $EW$. }
\end{figure}

Constant $k_0$ surfaces, which are spheres for $ PW $ $ (k_{\rho}^2+k_{z}^2=k_0^2) $ and hyperboloids for $ EW $ $ (k_{\rho}^2-\beta_{z}^2=k_0^2) $, are shown in Fig. \ref{fig:phasespacepwew}a and Fig. \ref{fig:phasespacepwew}b respectively (only $ k_z, \beta_z > 0 $ portion is shown).  
Since the expression for energy density (or stress tensor) is an integral over the in-plane wave vectors and frequency, we can identify $ \mu $ as the space spanned by $\{k_{\rho} \in (0,\infty),\phi \in (0,2\pi),k_0 \in (0,\infty)\}$ ($k_0$ is used instead of $\omega$). Consider a differential patch on a constant energy surface  (blue patch in Fig. \ref{fig:phasespacepwew}a and Fig. \ref{fig:phasespacepwew}b) whose projected area in the $ k_x-k_y $ plane is $ dk_xdk_y $ or $ k_{\rho}dk_{\rho} d\phi $. The area of this blue patch is given by $\displaystyle dS = \frac{k_{\rho} dk_{\rho} d\phi}{|\cos\theta_z(\mu)|} $, where $ \theta_z(\mu) $ is the angle between the surface normal to the constant energy surface and the $ k_z $ or $ \beta_z $ axis. $ \cos\theta_z(\mu) = \displaystyle k_{z0}/k_0 $ for $ PW $ and  $\cos\theta_z(\mu) = \displaystyle  \beta_{z0}/\sqrt{\beta_{z0}^2+k_{\rho}^2} $ for $ EW $.  $ d\mu $ is an infinitesimal volume element of 
area $ dS $ on a constant $k_0$ surface and thickness $ dk_0 $ perpendicular to it. The magnitude of $ d\mu $ is $\displaystyle  dk_0 dS$. 
 $ \int_{PW}d\mu$ and $ \int_{EW}d\mu$ represent triple integrals over the domains $\{\phi \in (0,2\pi), k_{\rho} \in (0,k_0), k_0 \in (0,\infty) \}$ and $\{\phi \in (0,2\pi), k_{\rho} \in (k_0,\infty), k_0 \in (0,\infty) \}$ respectively.   

Since $ \rho_a^{(j)}(\mu, z) $ is related to $ \rho^{(j)}(\mu, z) $, the integral expression for $ \rho(\omega, z) $ will be used to first determine $ \rho^{(j)}(\mu, z) $. 
In vacuum $ \rho(\omega,\mathbf{r}) =   \frac{\omega}{\pi c^2} \Im Tr (\overline{\overline{\mathbf{G}}}_e(\mathbf{r},\mathbf{r};\omega) + \overline{\overline{\mathbf{G}}}_m(\mathbf{r},\mathbf{r};\omega) )$ \cite{narayanaswamy2010a,*joulain2003definition}, where $ Tr(\overline{\overline{\mathbf{G}}}) $ is the trace of $ \overline{\overline{\mathbf{G}}} $, and $ \overline{\overline{\mathbf{G}}}_e(\mathbf{r},\mathbf{r};\omega), \overline{\overline{\mathbf{G}}}_m(\mathbf{r},\mathbf{r};\omega) $ are the electric and magnetic DGFs. $ \rho^{(j)}(\omega,z)d\omega $ \footnote{Unlike thermal energy density $u^{(j)}(z)$, it does not make sense to talk of $\int_0^{\infty} d\omega \rho^{(j)}(\omega,z)  $ since it is infinite} is given by:
\begin{equation}
\label{eqn:LDOS} \rho^{(j)}(\omega,z) d\omega= \int_{PW} ' d\mu  \rho^{(j)}(\mu, z) + \int_{EW}' d\mu  \rho^{(j)}(\mu, z) 
\end{equation}
where $\int_{PW} ' d\mu \equiv dk_0 \int_{PW} dS $ and $\int_{EW} ' d\mu \equiv dk_0\int_{EW} dS$ are double integrals over the domains   $\{\phi \in (0,2\pi), k_{\rho} \in (0,k_0) \}$ and $\{\phi \in (0,2\pi), k_{\rho} \in (k_0,\infty) \}$ respectively, and $\rho^{(j)}(\mu,z)$ is (Eq. \ref{eqn:rhojPW} for PW and Eq. \ref{eqn:rhojEW} for EW):
\begin{subequations}  
\label{eqn:rhoj} 
\begin{equation}
\label{eqn:rhojPW}
\begin{split}
\rho^{(j)}(\mu,z) = & \Re\Bigg\{\frac{1/4\pi^3}{(1-\tilde{R}^{(j)}_{0L}\tilde{R}^{(j)}_{0R} e^{i2k_{z0}l_0})} \times \\
 & \Bigg[\begin{array}{l}1+\tilde{R}^{(j)}_{0L}\tilde{R}^{(j)}_{0R} e^{i2k_{z0}l_0} + \frac{k_{\rho}^2}{k_0^2} \tilde{R}^{(j)}_{0R} \times\\  
 e^{i2k_{z0} (z_R-z)}+ \frac{k_{\rho}^2}{k_0^2} \tilde{R}^{(j)}_{0L} e^{i2k_{z0} (z-z_L)} \end{array} \Bigg] \Bigg\}
\end{split}
\end{equation}
\begin{equation}
\label{eqn:rhojEW}
\begin{split}
\rho^{(j)}(\mu,z) & =  \Im\Bigg\{\frac{1/4\pi^3}{(1-\tilde{R}^{(j)}_{0L}\tilde{R}^{(j)}_{0R} e^{-2\beta_{z0}l_0})\sqrt{2\frac{k_{\rho}^2}{k_0^2}-1}}  \\
 \times & \Bigg[\begin{array}{l}1+\tilde{R}^{(j)}_{0L}\tilde{R}^{(j)}_{0R} e^{-2\beta_{z0}l_0} + \frac{k_{\rho}^2}{k_0^2} \tilde{R}^{(j)}_{0R} \times\\  
 e^{-2\beta_{z0} (z_R-z)}+ \frac{k_{\rho}^2}{k_0^2} \tilde{R}^{(j)}_{0L} e^{-2\beta_{z0} (z-z_L)} \end{array} \Bigg] \Bigg\}
\end{split}
\end{equation}
\end{subequations}
From the dispersion relation, we see that $k_{z0} = \pm \sqrt{k_0^2 - k_{\rho}^2}$ and $\beta_{z0} = \pm \sqrt{k_{\rho}^2-k_0^2}$. Only the positive sign for $k_{z0}$ and $\beta_{z0}$ is valid in the expressions for $n^{(j)}_L(\mu,z)$ (Eq. \ref{eqn:njL}). In contrast, both signs of the square roots are valid in Eq. \ref{eqn:rhoj} for $\rho^{(j)}(\mu,z)$. 
 Using the property that $\tilde{R}_{0L}(-k_{z0}) = \tilde{R}_{0L}^{-1}(k_{z0})$ and  $\tilde{R}_{0R}(-k_{z0}) = \tilde{R}_{0R}^{-1}(k_{z0})$ \cite[see Eq. 2.7.3]{chew95a}, it can be seen that the same value of $\rho^{(j)}(\mu,z)d\mu$ is obtained irrespective of the choice of sign for the square root, i.e., $ k_z,\beta_z>0 $ and $ k_z,\beta_z < 0 $ portions of the $\mu$ space, corresponding to states accessible to waves originating from $L$ and $ R $ respectively, contribute equally to $\rho^{(j)}(\mu,z)d\mu$. Hence, for thermal non-equilibrium condition, $\rho_a^{(j)}(\mu,z) = \rho^{(j)}(\mu,z)/2$. The case of thermal equilibrium is discussed later.

The polarization dependent energy flux in the $ z $ direction due to sources in $ L $, $ \dot{E}^{(j)}_{L} $ (it is independent of $ z $), can be written as 
\begin{equation}
\label{eqn:flux}
\dot{E}^{(j)}_L=\int d\mu \cos \theta_z(\mu) \Theta(\omega,T_L) K^{(j)}(\mu)
\end{equation}
where $\Theta(\omega,T_L)=\hbar\omega/(\exp(\hbar\omega/k_B T_L)-1)$, and
\begin{equation}
\label{eqn:generalizedintensity}
\begin{split}
 K^{(j)}(\mu)  = \begin{cases}
\displaystyle \frac{c}{8\pi^3}   \frac{(1-|\tilde{R}^{(j)}_{0L}|^2) (1-|\tilde{R}^{(j)}_{0R}|^2)}{|1-\tilde{R}^{(j)}_{0L}\tilde{R}^{(j)}_{0R} e^{i2k_{z0}l_0} |^2}   \text{, } PW\\ 
\displaystyle \frac{c}{8\pi^3}    \frac{4\Im(\tilde{R}^{(j)}_{0L}) \Im( \tilde{R}^{(j)}_{0R}) e^{-2\beta_{z0} l_0}}{|1-\tilde{R}^{(j)}_{0L}\tilde{R}^{(j)}_{0R} e^{-2\beta_{z0}l_0} |^2}  \text{, } EW
\end{cases}
\end{split}
\end{equation} 
Biehs et. al. \cite{biehs2010mesoscopic} and Ben-Abdallah and Joulain \cite{ben2010fundamental} have interpreted $( 8\pi^3/c) K^{(j)}(\mu)  $ as a generalized transmissivity between the two half-spaces $ L $ and $ R $. Here, we interpret $ (\omega^2/c^3) \Theta(\omega,T_L) K^{(j)}(\mu)  $ as a \textit{generalized spectral radiation intensity}, valid for $ PW $ as well as $ EW $. The concept of radiation intensity ($ I $) in classical theory of thermal radiation is associated with the power contained in a cone of $ d\Omega $. For $ PW $, $ dS/k_0^2 $ can be associated with the solid angle in the direction of propagation because the constant $ k_0 $ surface happens to be spherical. The solid angle interpretation of intensity is invalid for EW. $ K^{(j)}(\mu)  $ on a constant $ k_0 $ surface for $ PW $ and $ EW $ (the entire surface is not shown) are shown in Fig. \ref{fig:phasespacepwew}c and Fig. \ref{fig:phasespacepwew}d respectively (simulation performed for $ s  $ polarization, $ l_0=5 $ $ \mu $m, $ 2\pi/k_0=10 $ $ \mu $m, $ \varepsilon_L = \varepsilon_R = 2.2+0.01i $). In each of the two figures, the length of the arrows are proportional to the magnitude of $ K^{(j)}(\mu)  $. The arrows are perpendicular to the constant $ k_0 $ surface. As $ k_{\rho} \rightarrow \infty $, the magnitude of the arrows decreases exponentially, in accordance with Eq. \ref{eqn:generalizedintensity}. 

$ \dot{E}^{(j)}_L $ can be also expressed as 
\begin{equation}
\label{eqn:energyfluxmuspace}
\dot{E}^{(j)}_{L}  = \int\limits d\mu  \hbar \omega n_L^{(j)}(\mu,z) v^{(j)}_{zL}(\mu,z)
\end{equation} 
 where $ v^{(j)}_{zL}(\mu,z) $ is the $ z $ component of the polarization dependent local velocity of energy transmission associated with photons from $L$. 
 Since  $ K^{(j)}(\mu) $ is the same for both $ L $ and $ R $ half-spaces, and $\cos\theta_z(\mu)$ is opposite in sign, $\dot{E}^{(j)}_{R}$ is opposite in direction to $\dot{E}^{(j)}_{L}$. At equilibrium, they cancel each other to yield zero net radiative transfer between the two half-spaces. Hence, at equilibrium $n_R^{(j)}(\mu,z) v^{(j)}_{zR}(\mu,z) = -n_L^{(j)}(\mu,z) v^{(j)}_{zL}(\mu,z)$.
 $ v^{(j)}_{zL}(\mu,z) $ can be shown to be:
\begin{equation}
\label{eqn:velocilty}
v^{(j)}_{zL}(\mu,z) =\begin{cases}
 \frac{c k_{z0}}{k_0}  \frac{  \frac{(1-|\tilde{R}^{(j)}_{0R}|^2)}{(1+|\tilde{R}^{(j)}_{0R}|^2)}}{1+\frac{k_{\rho}^2}{k_0^2}\frac{2\Re(\tilde{R}^{(j)}_{0R} e^{i2k_{z0}(z_R-z)})}{1+|\tilde{R}^{(j)}_{0R} |^2}}, \text{ }  PW
\\ 
  \frac{\frac{c \beta_{z0} k_0}{k_{\rho}^2}  \frac{2\Im(\tilde{R}^{(j)}_{0R})e^{-2\beta_{z0}(z_R-z)}}{(1+e^{-4\beta_{z0}(z_R-z)}|\tilde{R}^{(j)}_{0R}|^2 )}}{1+\frac{k_0^2}{k^2_{\rho}}\frac{2\Re(\tilde{R}^{(j)}_{0R})e^{-2\beta_{z0}(z_R-z)}}{(1+e^{-4\beta_{z0}(z_R-z)}|\tilde{R}^{(j)}_{0R}|^2 )}}, \text{ }   EW
\end{cases}
\end{equation} 
by comparing Eq. \ref{eqn:njL} and Eq. \ref{eqn:generalizedintensity}. 
We have confirmed numerically that, as one would expect, $ v^{(j)}_{zL}(\mu,z) \leq c$ for all $ \mu $. 
 The entropy flux, $ \dot{S}_{L}(z) $, associated with this energy flux is given by 
\begin{equation}
\label{eqn:entropyfluxmuspace}
\dot{S}_L(z)=\int d \mu \sum_{j=s,p}  s_{L}^{(j)}(\mu,z) v_{zL}^{(j)}(\mu,z)
\end{equation} 
$\dot{E}_R(z)$ and $ \dot{S}_R(z) $ can be obtained by the same procedure if $ T_L \neq T_R $. Unlike energy flux within the cavity which is independent of $ z $, though any two waves with different $ \mu $ are incoherent, interference effects due to multiple reflections of the same wave lead to a $ z$ dependence of entropy flux \cite{barakat1983nfold}. 

\begin{figure}
\centering
\includegraphics[width=8.0cm]{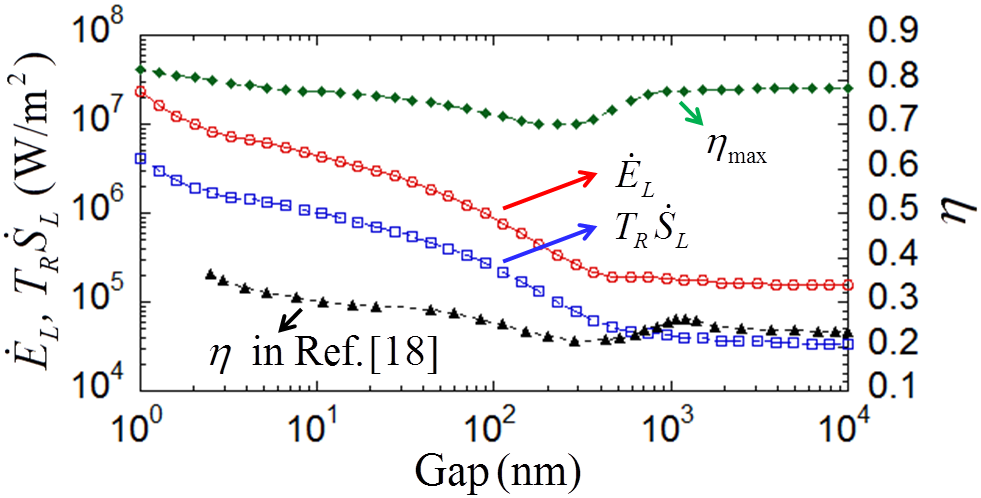}
\caption{\label{fig:Efficiency} (Color online) Energy and entropy flux (left $y$ axis, hollow markers) and conversion efficiency (right $y$ axis, solid markers) as a function of vacuum gap for W-In$_{0.18}$Ga$_{0.82}$Sb, compared to results in Ref. \cite{park2008performance}.}
\end{figure}

The maximum work that can be extracted from NFRT between the two half-spaces (rejecting radiation at sink temperature $ T_R $) in Fig. \ref{fig:config} is given by $ \dot{W}_{max} = (\dot{E}_L - T_R\dot{S}_L) - (\dot{E}_R - T_R\dot{S}_R) $ \cite{landsberg80a}. In this expression, both $ \dot{S}_L $ and $ \dot{S}_R $ are evaluated at $ z=z_R $. The maximum efficiency of work extraction is then given by $ \eta_{max} = \dot{W}_{max}/\dot{E}_L $. Park et. al. \cite{park2008performance} analyzed the performance of a thermophotovoltaic (TPV) converter using W emitter at 2000 K and In$_{0.18} $Ga$_{0.82} $Sb photovoltaic cell at 300 K. Using  optical data for W \cite{palik1} and In$_{0.18} $Ga$_{0.82} $Sb \cite{gonzalez2006modeling}, we compute $ \dot{E}_L,  \dot{S}_L, \dot{E}_R,$ and $  \dot{S}_R$, of which $ \dot{E}_L$ and $  T_R \dot{S}_L $ are shown in Fig. \ref{fig:Efficiency}, when $L$ is W at 2000 K and $R$ is In$_{0.18} $Ga$_{0.82} $Sb at 300 K. Both energy flux and entropy flux increase with decreasing gap, characteristic of tunneling due to evanescent waves. We also plot $ \eta_{max} $ as a function of gap and compare with the efficiency predicted by Park et. al.  \cite{park2008performance}. 
As observed by Landsberg and Tonge \cite{landsberg80a}, though the high values of $ \eta_{max} $ are usually unattainable, the utility of $ \eta_{max} $ is to impose an upper limit on efficiencies of all models for energy conversion, including TPV conversion, involving the same materials, configurations, and temperatures. 

The case of $ T_L=T_R = T$ has to be treated differently since emission from both half-spaces are at the same temperature. 
 When $ T_L=T_R$, electromagnetic waves in the vacuum cavity cannot be distinguished as originating from $ L $ or $ R $ because their temperatures are equal. Hence, in thermal equilibrium, it is $ n_L^{(j)}(\mu,z) + n_R^{(j)}(\mu,z) $ rather than $ n_L^{(j)}(\mu,z) $ or $ n_R^{(j)}(\mu,z) $ that determines the entropy. At the same time, $\rho_a^{(j)}(\mu,z)=\rho^{(j)}(\mu,z)$ since $k_z,\beta_z \geq 0$ and $k_z,\beta_z \leq 0$ portions of the dispersion surface are accessible now. 
 Since $ \frac{\sum\limits_{h=L,R}\hbar\omega n_h^{(j)}(\mu,z)}{\rho^{(j)}(\mu,z)} = \Theta(\omega,T)  $ at thermal equilibrium, the contribution to entropy density is given by $  \rho^{(j)}(\mu,z) \gamma(\omega,T)  $, where  $  \gamma(\omega,T)  = k_B \big[ \frac{\hbar \omega/k_BT}{\exp(\hbar \omega/k_BT)-1}-\ln[1-\exp(-\frac{\hbar \omega}{k_BT})]  \big]  $ \cite{planck}. Of this, a fraction $  \frac{\hbar \omega n_L^{(j)}(\mu,z)}{\Theta(\omega,T) \rho^{(j)}(\mu,z)} $ is from half-space $ L $ and the remainder from $ R $. The equilibrium entropy flux in the cavity due to thermal sources within $ L $ is given by $\dot{S}_L^{eq} = \int d\mu\sum\limits_{j=s,p}\hbar\omega  n_L^{(j)}(\mu,z) v_{zL}^{(j)}(\mu,z) \frac{\gamma(\omega,T)}{\Theta(\omega,T)}$ ($\dot{S}_L^{eq}$ is independent of $z$). Since $n_R^{(j)}(\mu,z)  v_{zR}^{(j)}(\mu,z)  = -n_L^{(j)}(\mu,z) v_{zL}^{(j)}(\mu,z)  $ at thermal equilibrium,   $ \dot{S}_L^{eq} $ and $ \dot{S}_R^{eq}  $ cancel each other and the net entropy flux is zero. 

In summary, we elucidate a method to determine entropy density and entropy flux between two half-spaces when near field effects are important. 
We also identified a generalized spectral radiation intensity which can be used even when NFE are present. Though the concepts developed here are used to analyze energy conversion using NFRT, it can also be used to better understand the thermodynamics of surface wave-based laser cooling \cite{khurgin2007surface} and thermal non-equilibrium  Casimir interactions \cite{kruger2011nonequilibrium,*Antezza05a,*Antezza06a}.
This work is funded partially by National Science Foundation Grant CBET-0853723.

%
\end{document}